\documentclass[useAMS,usenatbib]{mn2e}

\title[Short-term variability of QSO's]{Short-term optical 
variability of high-redshift QSO's}

\author[Bachev et al.]
	{R. Bachev$^{1,2}$, A. Strigachev$^{1}$, E. Semkov$^{1}$\\ 
$^{1}$Institute of Astronomy, Sofia 1784, Bulgaria; bachevr@astro.bas.bg, anton@astro.bas.bg;\\
$^{2}$Department of Physics and Astronomy, University of Alabama, Tuscaloosa, AL 35487, USA}

\date{}

\pagerange{\pageref{firstpage}--\pageref{lastpage}} 
\pubyear{2004}

\begin{document}

\maketitle

\label{firstpage}

\begin{abstract}
This paper presents results of a search for short-term variability in the optical band 
of selected high-luminosity, high-redshift radio-quiet quasars. Each quasar has been 
monitored typically for 2 -- 4 hours with a time resolution of 2 -- 5 minutes and a 
photometric accuracy of about $0\fm01$ -- $0\fm02$. Due to the significant redshift 
($z>2$), the covered wavelength range falls into the UV region (typically 1500 -- 
2500\AA). We found no statistical evidence for any continuum variations larger than 
$0\fm01$ -- $0\fm02$ for any of the monitored objects. Our results suggest that 
the presence of a short-term variability in radio-quiet quasars is unlikely even 
in the UV region, contrary to reports by other authors. This conclusion holds 
true at least for high-luminosity (large black hole mass and accretion rate?) 
objects. The results are consistent with the idea that significant short-term 
(less than 1 hour) variations in AGN, where observed, should be attributed primarily 
to processes in a relativistic jet.
\end{abstract}


\section{Introduction}

Optical continuum variability is one of the most distinct and definitive 
characteristics of quasars, revealed practically with the discovery of these 
objects (see Ulrich et al. 1997, for a review). It has been extensively 
studied because of its generally unknown nature and eventually because it can 
be used to restrain the models of the energetic processes that may take place in 
quasars and produce the continuum -- e. g. accretion, jet generations, etc. 
Here the shortest time-scale of the variability is of particular interest 
since it can impose an upper limit of the size of the emitting region. 
For instance, the X-ray continuum of some objects is known to vary
significantly on time-scales of minutes. Many authors have also 
found variations in the optical band on similar time-scales but 
smaller amplitudes (intra-night or micro-variability). 
For the radio-loud quasars, and especially the beamed ones
(i.e. blazar-type objects), such an intra-night variability 
is indeed a well-documented feature and is due presumably to rapid processes 
in a relativistic jet, invoking shock waves (Wagner \& Witzel 1995), 
plasma processes (Baker et al. 1988; Krishan \& Wiita 1994), etc. 
Interestingly however, some authors found such short-term 
variability with an amplitude of up to $\sim0\fm1$ in radio-quiet objects, 
where no jet contribution to the optical continuum is expected. 

Earlier microvariability observations were performed on relatively 
small-aperture telescopes, equipped with single-channel photo-multipliers. 
They supposedly revealed short-term variability of about $0\fm1$ in nearby 
Seyferts, monitored mainly in $U$ and $B$-bands (Lyutyi et al. 1989; 
Dultzin-Hacyan et al. 1992; Merkulova 2000). Unfortunately, with such an 
equipment an adequate assessment of the photometric errors is hardly 
possible and they, taking into account the photon noise, atmospheric 
instabilities, etc., may easily reach a value comparable to the 
variability amplitude claimed (see also the Discussion). 
Petrucci et al. (1999), on the other hand, monitored a large sample of
Seyfert galaxies in $I$-band with a CCD (i.e. a multi-channel photometric 
device) and did not find any evidence for microvariability down to the 
$0\fm01$ -- $0\fm02$ level. Similar results were reported by 
Gopal-Krishna et al. (1995), Sagar et al. (1996), Gopal-Krishna 
et al. (2000), Gopal-Krishna et al. (2003); hereafter GK03, 
Stalin et al. (2003); hereafter S03, Stalin et al. (2004), who 
generally did not find short-term variations above the $0\fm01$ -- $0\fm02$ 
level (mainly $R$-band) in a large sample of relatively luminous 
radio-quiet quasars (RQQ's) among other objects. These authors, however,
are convinced in the reality of smaller-amplitude variations, see \S 4 for 
a discussion. Several other CCD-based monitoring campaigns did not 
present convincing evidence for short-term variations in RQQ's: 
Rabbette et al. (1998) did not find variations above $\sim0\fm1$ in selected 
higher-$z$ RQQ's. Webb \& Malkan (2000) set an upper limit of the short-term 
variations of $0\fm03$ on time-scales of an hour or so for a sample of 23 
quasars (some radio-loud). Klimek et al. (2004) found only "signs" of 
short-term variations in a sample of narrow-line Seyfert 1 (NLS1) galaxies. 
Miller et al. (2000), on the other hand, presented evidence for short-term 
variations in an unusual NLS1 object, IRAS 13224-3809. 
The authors, however, do not show the temporal 
behaviour of a similar check star, so one cannot assess independently 
the real accuracy of their photometry. 
Similar short-term variation are reported in a nearby Seyfert 1 
galaxy Akn 120 (Carini et al. 2003), while several other similar 
objects showed no evidence of short-term variations. 


Based on the current literature on the subject one concludes that,
although generally a controversial issue, the short-term variability 
of RQQ's is either very rare or non-existing; most of the higher 
accuracy CCD monitoring campaigns simply do not find convincing evidence 
for it. In spite of the lack of evidence however, it should be pointed out that 
it is not entirely unexpected to find short-term variations in radio-quiet 
objects, particularly because the highly variable X-ray emission may have 
also an optical counterpart (Wiita 1996). If so -- a strong wavelength dependence 
should be expected in a sense that the microvariability will be much stronger at 
shorter wavelengths, taking into account the much higher temperature of the 
X-ray producing region (the central parts of an accretion disc or a hot 
corona above the disc). The exact wavelength dependence of the long-term 
variability has been debated for a long time but the latest studies seem 
to confirm that the variations increase with the wavelength decrease
(Di Clemente et al. 1996; Cristiani et al., 1997; Giveon et al. 1999; 
Trevese \& Vagnetti 2002; Ivezi\'{c} et al. 2004). It is not known for sure 
whether the nature of the long and short-term variations is the same, 
therefore it cannot be ruled out that the increase of the short-term 
variations toward the shortest wavelengths is even greater. 
Edelson et al. (1996) and Collier et al. (2001), based on multiwavelength 
monitoring of nearby Seyferts confirm that the short-term variations are 
indeed chromatic -- the far-UV continuum varies
stronger than the red continuum.

The assumption that the fast variations depend strongly on the wavelength 
can partially explain the apparent discrepancy in the results from 
the earlier observations in $U$ and $B$-bands and the later -- in $R$ and 
$I$-bands, in case that no instrumental effects or improper error 
handling is involved in either case. An ultimate test will be a 
high-accuracy CCD monitoring in UV (and far-UV) region.
For high-redshift quasars, this far-UV region is shifted to the 
optical band and therefore is easily accessible from earth.
Thus, for z$\sim$2 the $R$-band monitoring covers the 
region of 2000 -- 2500 \AA\ in the quasar's rest-frame and is 
practically a far-UV monitoring. 

In this paper we present the results of such a monitoring of 18 
high-redshift, high-luminosity RQQ's with an average redshift of 2.5 
and a typical visual magnitude of 16.5. Our goal is to clarify 
the role of the wavelength in the short-term variability as well 
as to fill the high-luminosity, high-redshift end in the 
distribution of QSO's, having been subjects of short-term variability 
monitoring campaigns so far.

\section{Observational data}
The observed objects (Tab. 1) were selected from the latest V\'{e}ron-Cetty \& 
V\'{e}ron (2003) catalog, hereafter VCV03, and are shown in Tab. 1. Col. 1 displays 
the most common name of the object (taken from VCV03). 
The redshift is shown in Col. 2. Col. 3, 4 and 5 display $VR_{\rm c}I_{\rm c}$ 
magnitudes of the quasars -- they are either taken from the literature or 
based on our measurements, accurate to $\sim 0\fm1$. The absolute $V$-band 
magnitude ($M_{\rm v}$) is calculated based on $V$ from Col. 3 ($q_{\rm 0}=0$ 
and $H=50$ kms$^{-1}$Mpc$^{-1}$ is adopted, as done in VCV03), and is shown 
in Col. 6. The last column (7) indicates the source for the magnitudes 
in Col. 3, 4 and 5 -- NED is the NASA/IPAC Extragalactic Database, 
VCV is VCV03, and t.w. stands for this work.

\subsection{Object selection}

\begin{table}
  \caption{Objects}
  \begin{tabular}{lrrrrrr}
  \hline
  \hline

Object Name &  z & V & $R_{\rm c}$ & $I_{\rm c}$ &
 $M_{\rm V}$ & Refs.\\
 (1) & (2)  & (3)  & (4)  & (5)  & (6)  & (7)  \\
 \hline

Q 0013+0213  &1.55&16.4&    &    &$-$29.0&NED  \\
S5 0014+81   &3.39&16.5&    &    &$-$31.1&NED  \\
PHL 957      &2.69&16.6&    &    &$-$30.3&VCV  \\
UM 673	 &2.72&17.0&16.8&    &$-$29.9&NED  \\
Q 0226$-$1024&2.27&16.3&15.5&14.6&$-$30.1&t.w. \\
HS 0741+4741 &3.20&16.4&16.2&16.1&$-$31.0&t.w. \\
PG 1247+268	 &2.04&15.6&    &    &$-$30.5&VCV  \\
HS 1312+7837 &2.00&15.8&    &    &$-$30.2&VCV  \\
SBS 1425+606 &3.16&15.8&    &    &$-$31.5&VCV  \\
SBS 1542+541 &2.37&17.1&16.9&    &$-$29.4&t.w. \\
HS 1603+3820 &2.51&16.2&15.9&    &$-$30.5&t.w. \\
HS 1626+6433 &2.32&16.5&16.3&    &$-$30.0&t.w. \\
HS 1700+6416 &2.74&16.1&15.7&    &$-$30.9&t.w. \\
SBSS 1711+579&3.00&18.0&    &    &$-$29.2&NED  \\
HS 1946+7658 &3.05&16.2&15.8&    &$-$31.5&NED  \\
HS 2103+1843 &2.21&16.8&    &    &$-$29.5&VCV  \\
HS 2140+2403 &2.17&17.8&    &    &$-$28.5&VCV  \\
HS 2337+1845 &2.62&16.9&16.7&    &$-$29.9&VCV  \\

\hline
\hline
\end{tabular}

\end{table}

\begin{table*}
 \centering
  \caption{Observations}
  \begin{tabular}{@{}llrrlrrrrrrrr@{}}
  \hline
  \hline

Object Name & Date & Inst. & Filt. & $\lambda_{\rm Rest}$ & $T_{\rm Tot}$ & 
$T_{\rm Rest}$ &  $t_{\rm Exp}$ & N & Aper. & $\sigma$ & $\Delta m$ & P(\%)\\
  (1) & (2)  & (3)  & (4)  & (5)  & (6)  & (7) & (8) & (9)&(10) &(11) &(12) &(13) \\
 \hline

Q 0013+0213    & 28.08.2003  & 0.6 AOB  & --  & 1770--3730  & 2.8 & 1.1 & 150  & 61 & 8 & 0.018 & 0.018 & 0 \\%
S5 0014+81     & 03.11.1999  & 0.6 AOB  & --  & 1030--2160  & 4.4 & 1.0 & 180  & 47 & 6 & 0.015 & 0.019 & 50 \\%
PHL 957        & 20.09.2002  & 1.3 SkO  & R  & 1520--2090  & 1.8 & 0.5 & 300  & 20 & 6 & 0.015 & 0.022 & 81 \\%
UM 673	   & 04.09.2003  & 1.3 SkO  & R  & 1500--2070  & 2.7 & 0.7 & 300  & 29 & 6 & 0.006 & 0.007 & 31\\%
Q 0226$-$1024  & 17.12.2003  & 0.6 AOB  & --  & 1380--2900  & 3.4 & 1.0 & 120  & 61 & 8 & 0.012 & 0.013 & 3\\%
HS 0741+4741   & 17.12.2003  & 0.6 AOB  & --  & 1070--2260  & 2.7 & 0.6 & 120  & 65 & 12 & 0.023 & 0.022 & 0\\%
PG 1247+268  & 21.03.2004 & 2.0 NAO & R & 1840--2530 & 4.9 & 1.6 & 300 & 47 & 5 & 0.003	 & 0.002 & 91\\%
HS 1312+7837 & 13.05.2004 & 0.5 NAO & -- & 1500--3170 &  3.9 & 1.3 & 300 & 28 & 7 &
0.037 & 0.034 & 13\\%
SBS 1425+606 & 12.05.2004 & 0.5 NAO & -- & 1080--2280 & 3.9 & 0.9 & 300 & 40 & 7 &
0.056 & 0.047 & 23\\%
SBS 1542+541   & 01.06.2003 & 0.6 AOB  & --  & 1340--2820  & 2.8 & 0.8 & 180  & 49 & 8 & 0.028 & 0.027 & 0\\%
HS 1603+3820   & 03.06.2003 & 0.6 AOB  & --  & 1280--2700  & 3.3 & 0.9 & 180  & 56 & 6 & 0.014 & 0.013 & 0\\%
HS 1626+6433   & 31.05.2003 & 0.6 AOB  & --  & 1350--2860  & 3.2 & 1.0 & 180  & 60 & 6 & 0.018 & 0.018 & 0\\%
HS 1700+6416   & 30.05.2003 & 0.6 AOB  & --  & 1200--2540  & 3.8 & 1.0 & 180  & 59 & 8 & 0.013 & 0.016 & 28\\%
SBSS1711+579  & 16.06.2002 & 0.6 AOB  & --  & 1130--2380  & 3.1 & 0.8 & 120  & 73 & 6 & 0.032 & 0.037 & 9\\%
HS 1946+7658   & 17.09.2002 & 1.3 SkO  & R  & 1380--1900  & 3.1 & 0.8 & 1800 & 7  & 6 & 0.003 & 0.002 & 5\\%
HS 2103+1843   & 17.09.2002 & 1.3 SkO  & R  & 1750--2400  & 1.4 & 0.4 & 300  & 16 & 6 & 0.009 & 0.010 & 39\\%
HS 2140+2403   & 12.08.2003 & 1.3 SkO  & R  & 1770--2430  & 4.2 & 1.3 & 300  & 45 & 6 & 0.012 & 0.017 & 85\\%
HS 2337+1845   & 16.09.2002 & 1.3 SkO  & R  & 1550--2130  & 3.8 & 1.0 & 300  & 31 & 6 & 0.011 & 0.015 & 58\\%

\hline
\hline
\end{tabular}
\end{table*}

The objects were selected upon several criteria (see below). Despite our best 
intentions, few objects do not satisfy one or another formal condition, 
although we include the results of their monitoring since they do not 
alter our main conclusions.

\begin{enumerate}

\item \textbf{Radio quietness}. 
All the objects were required to be radio-quiet or 
to have undetected radio emission, which at least suggests that they are not 
significantly radio-loud in terms of the Kellermann index 
($R_{\rm K}=L_{\rm Radio}/L_{\rm Opt}$), Kellermann et al. (1989). 
This is the most important requirement since we want to isolate the 
short-term variability due to a jet from the one that could possibly be 
associated with an accretion disc. The only exception is S5 0014+81 which 
has $R_{\rm K}\simeq250$ and technically should be considered radio-loud. 
Any possible detection of micro-variations of this object should be 
interpreted with care.

\item \textbf{Photometric errors $\sigma \simeq 0\fm01 - 0\fm02$ and a time 
resolution of the light curves of 5 minutes or better}. 
This requirement is important for a successful time and amplitude 
resolution of the micro-variations as they are expected to occur in AGN's. 
It naturally means visually bright objects. To satisfy this 
requirement by improving the signal-to-noise ratio of the photometry, some 
objects had to be observed with no photometric filter applied (see \S 2.2). 
Still, due to unexpected weakness of some objects or non-excellent atmospheric 
conditions, this condition could not be entirely satisfied for some objects 
(Tab. 2).

\item \textbf{High redshift -- $z>2$}. 
This requirement is imposed in order to insure that the far UV-region is 
observed, even in cases where no filter is used (see below). Thus, the 
covered rest-frame wavelengths region falls generally into the 1000 -- 3000 \AA\ 
range (Tab. 2). This requirement was fulfilled for all the objects, 
except for Q 0013+0213 ($z=1.55$). 

\item \textbf{Convenient comparison stars in the field}. 
We required at least one bright star (main standard) to be  
in the field of the quasar and at least one more check star of 
moderate brightness, which is essential for a reliable differential 
photometry. This requirement was easily satisfied for all 
QSO's we initially intended to observe. 

\end{enumerate}

Several tens of quasars from VCV03 satisfy the requirements above. We naturally had 
to exclude objects with a declination $\delta < -15\degr$ (see the next section).

\subsection{Observations and photometry}

The observations were performed with the 0.6-m reflector of 
Belogradchik Observatory, Bulgaria, equipped with a CCD SBIG ST-8 camera 
(Bachev et al. 1999), the 2.0-m Rozhen National Observatory, Bulgaria, 
equipped with a Photometrix AT200 CCD, the 0.5/0.7-m Schmidt camera of the 
same observatory, equipped with a SBIG ST-8 and the 1.3-m reflector of 
Skinakas Observatory, Crete, where a Photometrix CH360 camera 
was used. The telescopes were equipped with standard 
$BVR_{\rm c}I_{\rm c}$ filters. All observations were performed 
in presumably clear, photometric nights. In order to diminish any 
possible atmospheric effects, the objects were preferably monitored 
during culmination, where the air mass does not change 
much for the time of the monitoring. This also insured a relatively 
good seeing -- typically 1.5 -- 2.5 arcsec. 

Tab. 2 presents the observations. Each object was observed once (Col. 2) 
with the instrument shown in Col. 3. Col. 4 shows the applied broad-band 
filter, if any. For the bigger Rozhen and Skinakas telescopes a standard 
$R_{\rm c}$ filter was used since it is there where the maximum of the 
camera's sensitivity is. For the smaller Belogradchik and Rozhen telescopes no 
filter was used. The spectral sensitivity of the camera in use there 
roughly mimics the transitivity of an $R$-filter (but is broader, of course; 
www.sbig.com). This approach helped to improve significantly the statistics -- 
by using of $R$-filter for instance we lose $\sim 1\fm3$, 
compared to frames taken with no filter. Since the covered rest-frame 
region falls into the far-UV region, it is understandable that the 
standard $BVR_{\rm c}I_{\rm c}$ bands do not have the same meaning 
as for zero-redshift objects. 
Therefore by not restricting the wavelength region (by not applying a filter) 
we do not alter our initial intention to monitor the UV-continuum 
in the quasars' rest-frame. Possible caveats are discussed in Sec. 4. 
The covered region is given in Col. 5 of Tab. 2; the limits indicate 
approximately the 25\%\ level of the maximum of the filter transmission 
or the camera sensibility respectively.   
The total monitored time in hours is given in Col. 6. Our best intention 
was to cover all the objects for about 4 hours. Unfortunately, this could not
be achieved for many objects, basically because of unstable 
atmospheric conditions during some nights. Note that because of the high 
redshift the total monitoring time is much shorter in the quasars restframe 
-- the rest-frame monitoring time is given in Col. 9 and is typically of 
about an hour or so. The frames' exposure time (in seconds) and the total 
number of exposures are given in Col. 8 and 9.

After dark subtraction and flat fielding, the magnitudes of the quasar and 
the comparison stars were extracted by applying standard aperture photometry 
routines (e.g. DAOPHOT under MIDAS, see also Bachev et al. 1999). 
The aperture diaphragms were taken to be typically 2 -- 3 times
the seeing. Several diaphragms were tested and the 
one giving minimal errors (as well as minimal standard deviation for the 
differential light curves between the comparison stars) was chosen. 
This diaphragm is given in Col. 10 in arcsec. Col. 11 gives the photometric 
error ($\sigma$) of the quasars' magnitudes. It is important to note that 
this is the formal, theoretical error, calculated based on the photon 
noise only (see the Discussion section for details). This error takes 
into account the formal error of the main standard as well, i.e. it is 
the formal error of the difference (QSO -- St1).

\section{Results}

Fig. 1 -- 3 show the results of our short-term variability search. 
The top panel of each box displays the magnitude difference between the 
quasar of interest and the main standard, the bottom one -- the difference 
between a check star and the main standard. 
The check star was chosen on a basis of spatial and magnitude 
proximity to the quasar. We preferred to use check 
stars of similar magnitude as the monitored object, since this is the best 
way to account for the real photometric errors of an object of such magnitude. 
As many authors pointed out, the formal (theoretical) errors that the 
program codes return, are usually smaller than the real errors by a factor 
of typically 1.5 -- 1.75 (S03, and the references within).
Our analysis indicates on average $\sigma_{\rm Real}/\sigma_{\rm Form}\simeq1.3$. 
This implies that the errors indicated in Table 2 for the quasars' 
photometry should be multiplied by $\sim 1.3$ in order to represent 
the real errors of photometry. Interestingly however, we found some 
signs for magnitude dependence of this ratio -- it is large for very 
small $\sigma$ and approaches 1 when $\sigma\ga0\fm02-0\fm03$. 
This behaviour is consistent with idea that the real photometric errors 
approach the theoretical when the photon noise is the primary error 
source and approach some minimum $>0$ when it is not. 
This minimum turns out to be of the order of $0\fm005$ -- $0\fm01$ 
and this should be considered as the accuracy limit of our differential 
photometry (Sect 4.2).  

Table 2 (Col. 12) shows the standard deviation ($\Delta m$) of the (QSO -- St1) 
light curve for the monitored period. For most of the objects this 
$\Delta m$ is statistically indistinguishable even from the theoretical 
photometric error (Col. 11). Col. 13 shows the results of a $\chi^2$ analysis, 
assuming that $\sigma=\sigma_{\rm Real}=1.3\sigma_{\rm Form}$, i.e. it gives 
the probability of ruling out the null hypothesis of non-variability. 
It is seen that from a statistical point of view, none of the objects 
monitored shows variability above the 95\%\ confidence level. 
In other words, the main result of our work indicates no
presence of short-term variability in any of the monitoring objects. 

\begin{figure}
 \mbox{} \vspace{8.5cm} \includegraphics{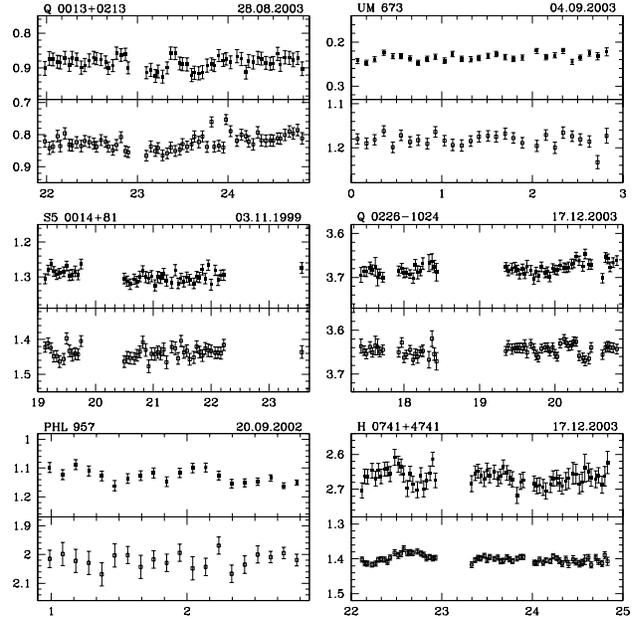} 

\caption[]{Results from the short-term variability search. The upper panel of each 
box shows the magnitude difference ($m_{\rm QSO}-m_{\rm St1}$) 
(filled simbols), the lower one -- ($m_{\rm St2}-m_{\rm St1}$) (open simbols), 
where St1 is the main standard and St2 is a check star of a magnitude as 
close as possible to the magnitude of the quasar (see the text). 
Theoretical (see \S 3) photometric errors at 1-$\sigma$ level are 
indicated as error bars. 
The name of each monitored object as well as the date of observations 
are shown on the top of each box. The abcissa represents UT in hours, 
i.e. one abcissa subdivision corresponds to 10 minutes. Each upper 
panel has the same vertical scale factor as the corresponding lower panel, 
but the value may differ from object to object.}

\end{figure}

\begin{figure}
 \mbox{} \vspace{8.5cm} \includegraphics{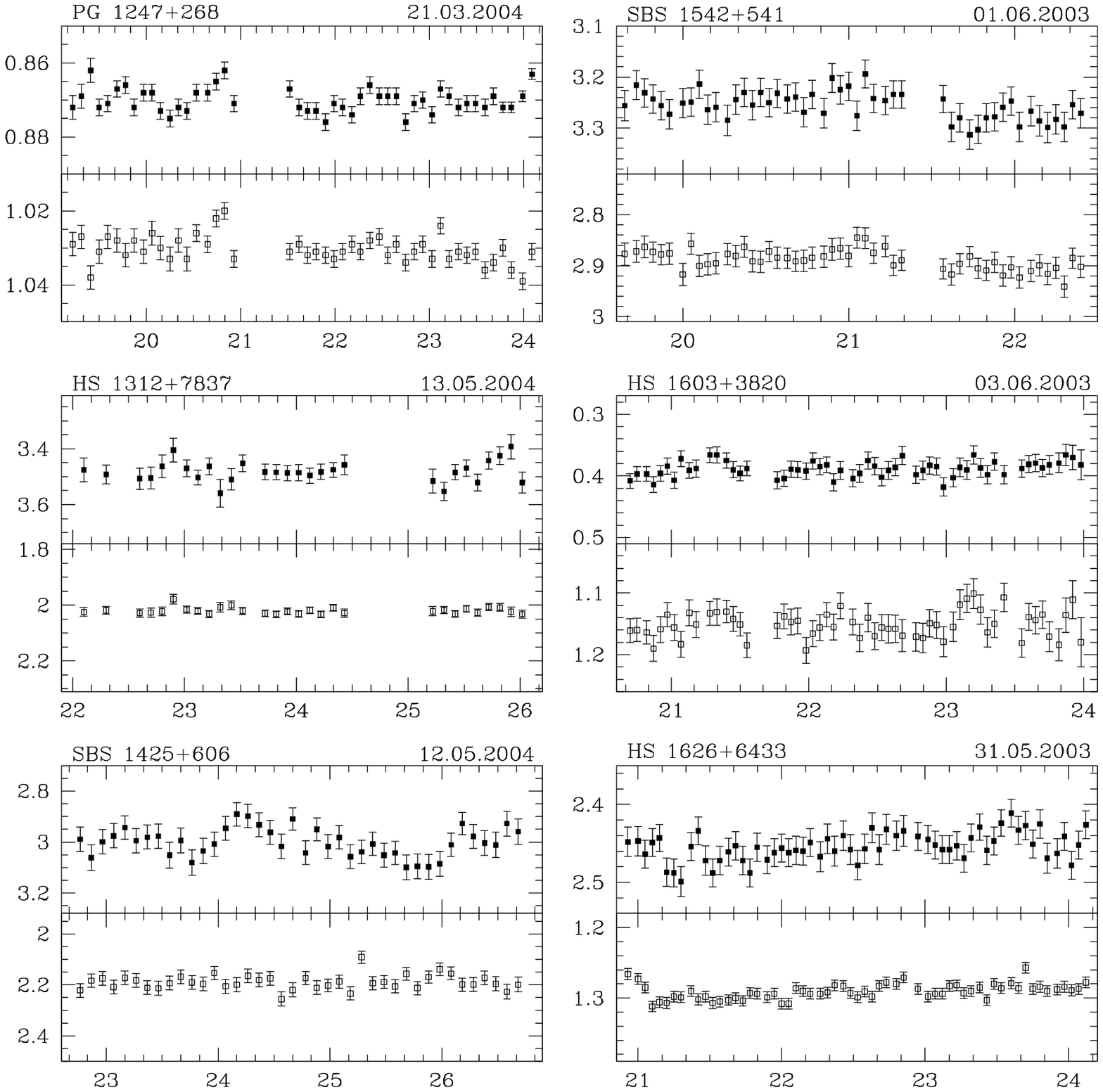} 
\caption[]{See Fig. 1 }
\end{figure}

\begin{figure}
 \mbox{} \vspace{8.5cm} \includegraphics{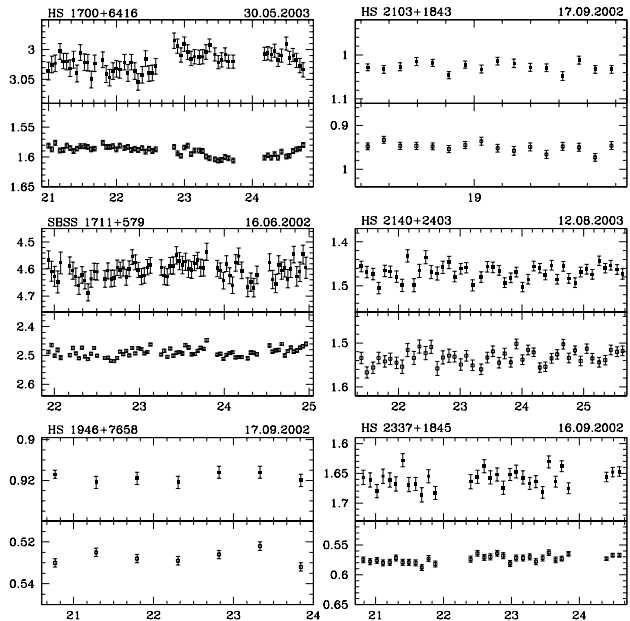} 
\caption[]{See Fig. 1 }
\end{figure}

\section{Discussion}

\subsection{High-$z$ caveats}
Monitoring high-redshift objects has the advantage that the far-UV region is 
shifted to the optical band, and therefore is directly accessible to the ground 
observations. Furthermore, there is no need of a broad-band filter to 
restrict the wavelength region covered, as far as the entire
UV continuum is assumed to be emitted by a relatively small region -- the inner 
portion of an accretion disc. As a disadvantage of the high $z$ can be 
mentioned the corresponding contraction of the total monitoring time in QSO's 
rest-frame (Tab. 2), which makes a monitoring of few hours probably not enough to 
reveal a possible variability (\S 4.3). For very high-$z$ objects ($z>4$)
the Lyman forests fall into the visual range, therefore such objects should
better be monitored in the IR.

When monitoring the UV continuum, one is to take into account that possible strong 
emission lines fall into the observed region, like Ly$\alpha$, CIV$\lambda1549$, 
FeII multiplets, etc. The line-emitting region is located further out from the central 
parts of an accretion disc and is not expected to be variable over intra-night 
time-scales. Therefore, the presence of such lines may lead to underestimation of 
the deducted continuum variability amplitude (or the upper limit of such) and appropriate
correction may be in order. This effect in principle is present in any 
spectral region but is especially strong in the UV; the equivalent width of the 
lines there can easily exceed 300 -- 400\AA, which means that the lines can contribute 
with up to 50\%\ to the total flux from the region. 

A sample of high red-shift quasars, due to the selection effects, by all 
means implies a sample of very luminous quasars (Tab. 1). 
Statistically, the highest luminosity, presumably not beamed quasars 
in the Universe, even if their luminosity is somewhat amplified 
by macrolensing, would most probably be powered by black holes of extremely large mass 
($M_{\rm BH}\simeq10^{9-10}$ M$_{\odot}$) fed at about the Eddington limit. Possible implications are discussed in \S 4.3.

\subsection{Photometric errors}
The proof of the existence of any short-term variations, especially the 
small-scale ones, depends critically on the correct assessment of the 
photometric errors. Since the atmosphere acts like a colour filter of variable 
transparency, the differential photometry of two stars of different colours 
should be affected by the air-mass changes during the monitoring. Especially 
prone to atmospheric effects seems to be the practice of not using a filter 
during the monitoring (\S 2). As S03, however pointed 
out, the effect of the colours of the comparison stars on the differential 
light curve is negligible for a specific band (see also Carini et al. 1992). 
Since the cameras we used (\S 2) are basically sensitive in $V$, $R$ and 
$I$-bands, not using a filter means that we add the signal in these three 
bands together; if any colour effect is negligible for an individual 
filter this should hold true for the sum as well. The effect ought to 
be explored carefully should the camera be highly sensitive 
in $U$ and $B$-bands, where extinction coefficients in reduction 
equations are the largest (see S03, for details). 
In any case, should we have observed any trends in the differential 
light curves, we would search for any magnitude -- air-mass correlation; 
this should be done any time when trends are observed independently 
whether a filter is used or not (e.g. Klimek et al. 2004).
 
Our analysis of the $\sigma_{\rm Real}/\sigma_{\rm Form}$ ratio 
suggests the presence of a minimal error of the photometry (see \S 3), 
which for our observations is estimated to be $\sim0\fm005$ but probably 
may vary. The presence of a minimal error may or may not be the case of other 
researchers' photometry -- in any case such a possibility should be 
carefully explored. It is understandable that if this is really the 
case, by adjusting the individual $\sigma_{\rm Form}$ by the average 
$<\sigma_{\rm Real}/\sigma_{\rm Form}>$, we overestimate the errors 
of some objects but underestimate others, what may lead to a wrong 
conclusion about the variability properties of the latter. 
In any case, having check stars of a magnitude, close to the one of 
the quasar, seems to be the correct way to account for any 
spurious sources of errors. 

One can think of many possibilities leading to an upper limit of the 
photometric accuracy. Non-perfect flat-field correction, parasite light 
and optics imperfections are among them. It is not that important to 
know the sources of all errors, what is actually important is to 
be aware of their magnitude.

\subsection{Comparison with other results}
Since some researchers find clear indications of short-term variability in RQ 
objects, an immediate question would be why our results do not reveal such. 
There are several possibilities, related to the following: 

\begin{enumerate} 

\item \textbf{Quasar parameters}

By short-term variability it is often assumed intra-night variability. 
However the objects observed may be very different -- i.e. variations 
that are well seen over a night in a low-mass NLS1 nucleus might require 
much longer time to be detected in a powerful quasar. In fact, most 
of the monitoring campaigns do not distinguish between quasars with 
different accretion parameters (e.g. black hole mass, accretion rate, 
etc.). As we mentioned above, our objects would probably have very 
large black hole masses, what may result in larger variability times. 
If the UV-continuum variations occur at a fixed distance (in 
Schwarzschild radii) from the black hole and reprocessing of hard radiation 
is primarily responsible for them, the variability time-scale will scale 
linearly with the mass (see also \S 4.4). The time contraction 
(due to the high redshift) additionally worsens the situation in our case.

Since the time-scale of the variability is related to the variability
amplitude (through the structure function, di Clemente et al. 1996),
not enough time coverage may equally mean not good enough photometric 
accuracy as a possible explanation of non-detection of the variability.
Indeed, it is often assumed that the long-term variability amplitude 
anticorrelates with the luminosity (Paltani \& Courvoisier 1997; 
Cristiani et al. 1997). The exact relation between the variability and 
luminosity, however, is a subject of discussions. Furthermore it may 
or may not be the same for the short-term variations.

\item \textbf{Interpretation of variability results}



When compare our results with results from other studies, where short-term 
variations in RQQ's are sometimes reported, one sees that often these studies 
have better accuracy and time coverage than our study. It is obvious that this 
may be one of the reasons why we did not detect any variations in our QSO sample. 
One however, should be somewhat sceptical when photometry of extremely high accuracy is reported. We already mentioned that our analysis suggests the presence of an upper 
limit of the photometric accuracy (\S 4.2). This effect ought to be explored carefully. 
The role of possible instrumental effects, which in general may be difficult to 
account for, also should not be underestimated. Klimek et al. (2004), for instance 
give a good example of how spurious effects, related to seeing, position of the star 
on the chip, etc., can affect the measured magnitudes. The exposure time can also 
affect $\sigma_{\rm Real}/\sigma_{\rm Form}$ ratio by the number of cosmic rays, for 
instance (those can hardly be fully removed). One indication of a possible 
underestimation of the real errors of the photometry is finding many "variable" 
comparison stars in quasar fields. It is statistically improbably that significant 
number of random stars will turn out to be variable on time-scales of an hour or so. 
Such comparison stars' variations are indeed reported in studies, where the quasar 
of interest is also sometimes found to vary similarly (Gopal-Krishna et al. 1995; 
Sagar et al. 1996; Gopal-Krishna et al. 2000; GK03; S03). 
In any case, for those objects, for which evidence for short-term 
variability is presented, an independent confirmation is by all means required.

The presence or absence of variability is usually assessed based on a 
statistical criterium. Different authors, however, often use different 
variability criteria. 
Computing quantities like the "amplitude of the variations" 
$A_{\rm m} = (\Delta m^{2} - \sigma_{m}^2)^{1/2}$, (Dultzin-Hacyan et 
al. 1992), the variability coefficient $C_{\rm eff}=\Delta m /\sigma_{m}$ 
(Jang \& Miller 1997; S03) or $P(\chi^{2})$ (Webb \& Malkan 
2000) constitute different approaches that may lead to different 
conclusions on the presence of the variability, even applied to the same data. 
This is true especially when the variations are very small or negligible.
We believe that any variability amplitude calculations should be performed
only after a statistical criterium, taking into account the number of the
observational points (e.g. $\chi^{2}$), proves undoubtedly the reality 
of the variations (by ruling out the non-variability hypothesis at 
significant level). Such an approach we follow in this paper.


\end{enumerate}

\subsection{Implications for variability models}

A non-detection of variations over a certain time scale in a class of 
objects is not necessarily a negative result, since it still provides 
some constraints on the physical models of the central engine. 

It is understandable that if some thermal mechanism is responsible 
for the continuum changes at certain wavelength $\lambda$ 
(far-UV in our case), these changes should mostly occur in the regions 
where emission of such $\lambda_{\rm UV}$ is primarily comming from, 
i.e. regions having the corresponding high enough temperature. It is so, 
because a temperature change affects in a greater extend the Rayleigh-Jeans 
part of the Planck curve, rather than the Wien part, i.e. the continuum around 
$\lambda_{\rm UV}$ will be effectively enhanced only if already
$T_{\rm Eff}(r) \ga \frac{hc}{k_{\rm B}\lambda_{\rm UV}}$. Since we here consider 
basically one spectral region, we can use the expression for the 
radial dependence of the temperature, assuming a standard Shakura-Sunyaev disc -- 
$T_{\rm Eff}(r) \propto \dot m^{1/4}M_{\rm BH}^{1/2}r^{-3/4}$ 
(e.g. Frank et al., 2002) to find the dependence of the radius of a 
fixed temperature on the accretion parameters, i. e.  
$r_{\rm UV} \propto \dot m^{1/3}M_{\rm BH}^{2/3}$. 
Here $\dot m$ is the accretion rate (in Eddington units), $M_{\rm BH}$ is the 
black hole mass and $r$ is the radius in linear units. Thus $r_{\rm UV}$ will be the 
radial distance of the region, primary responsible for the UV-continuum changes; 
further out the temperature is too low; inward the area decreases (and
respectively the total flux).

The shortest variability time-scale ($\tau_{\rm var}$) is usually associated 
with the light-crossing time and implies reprocessing of hard (X-ray) 
radiation into UV-optical region (Ulrich et al. 1997).
If the variable hard X-ray emission originates in the centre, then 
$\tau_{\rm var}^{\rm (hours)} \simeq 0.25 M_{\rm 8}(r_{\rm UV}/R_{\rm S})$
or $\tau_{\rm var} \propto r_{\rm UV}\propto \dot m^{1/3}M_{\rm BH}^{2/3}$.  
Here $M_{\rm 8}$ is the central mass in $10^{8}$ M$_{\odot}$ and $R_{\rm S}$ 
is the Schwarzschild radius. One sees, therefore (\S 4.3), 
that this time may not be enough to reveal any intra-night variations 
in a powerful quasar ($M_{\rm BH}\simeq10^{9-10}$ M$_{\odot}$).
On the other hand, the variable hard radiation may not necessarily come 
from the centre, but from a hot corona above the disc instead 
(Merloni \& Fabian 2001). The resulting $\tau_{\rm var}$ in such a case can 
be much shorter. 
One can explore roughly the amplitude dependence on the accretion parameter
based on a very simple model: If the size of an active region 
producing short-term UV flares is fixed to $l$, 
the variability amplitude (assuming random flare-producing events 
of number $N$ above a flat reprocessor -- accretion disc) should be
estimated as $N^{-1/2} \propto l/r$ or respectively 
$\propto \dot m^{-1/3}M_{\rm BH}^{-2/3}$. 
This relation, for an Eddington-limited accretion, gives $\sim100$ 
times smaller amplitude of the microvariations for a powerful quasar, 
compared to a NLS1 galaxy with $M_{\rm BH}\simeq10^{6}$ M$_{\odot}$.
In both cases we see that it is far more likely to observe short-term
variations in AGN of lower central mass (e.g. NLS1, mini-Seyferts).

If the accretion operates through an optically thick advective disc 
(i. e. slim disc; Abramowicz et al. 1988), the radial temperature scales 
as $T_{\rm Eff}(r)\propto M_{\rm BH}^{1/2}r^{-1/2}$ (Watarai \& Fukue 1999). 
Similar analysis will reveal in such a case that the amplitude of the variations 
scales as $M_{\rm BH}^{-1/2}$.



Another possible explanation of the variations is based on the instabilities of 
the flow itself, rather than on reprocessing. If assume that the inner part of 
the flow operates through an optically thin advective mode (ADAF; Narayan et al. 
1998), the border between the ADAF and the outer thin disc may be a good 
candidate for the region where these instabilities occur 
(Gracia et al. 2003; Krishan et al. 2003). Again, far-UV 
variations will be effectively generated if this transition happens to 
occur close to $r_{\rm UV}$ (see above).
The time-scale of such changes may be associated with the accretion time-scale 
of the ADAF (close to the free-fall time-scale), which is about 
$\tau_{\rm var}^{\rm (hours)} \simeq 0.4M_{\rm 8}(r/R_{\rm S})^{3/2}$ and may be 
as small as few hours in case of very small ADAF section. 
The thin disc -- ADAF transition depends on the accretion parameters and 
is supposed occur at about the last stable orbit if the accretion rate is relatively
high -- $\dot m \ga 0.1$, (R\'{o}\.{z}a\'{n}ska  \& Czerny 2000). 
Yet, for a very massive and powerful quasar such a mechanism does not seem to 
be able to generate significant short-term variations.  

If any variations are detected, the variability pattern (the structure function) 
will possibly be able to distinguish between explosive events and instabilities 
at the ADAF -- thin disc border. 

\section{Conclusions and Summary}

In this paper we present results from a short-term variability monitoring
of luminous high-redshift radio-quiet QSO's. The far-UV (restframe) continuum
has been observed for several hours (about an hour in the restframe). No
statistical evidence for variations above the $0\fm01$ -- $0\fm02$ level has been 
found in any of the observed objects. We conclude that such variations, if existing 
at all, are untypical at least for these objects. On the other hand, it is completely 
possible that our photometric accuracy and time coverage may simply not be 
enough to reveal possible variations at the micro level, since such are sometimes 
reported in other studies. We stress, however, on the importance of the correct 
assessment of the photometric errors for the interpretation of the variability 
results, suggesting that in some of these studies the errors may have been 
underestimated.


In spite we failed to detect any short-term variations in a large sample of quasars,
our analysis suggests that such may be well detectable in some objects with smaller black 
hole masses (e.g. NLS1 or mini-Seyferts). This possibility is supported by the 
fact that direct (space) observations of the UV continuum indeed reveal day-scale 
variations in some objects. Based on our experience and taking into account the 
results of other researchers, we can propose several possible further steps that 
can be successfully performed on small telescopes, and contribute significantly 
for the understanding of the nature of the short-term variability in QSO's:

\begin{enumerate} 
\item Observations of high-redshift luminous QSO's on time scales of days or 
weeks in order to reveal the shortest variability time-scale of these powerful
objects.

\item Observations of a selected lower-redshift NLS1's to clarify the role of 
the mass and accretion rate on the variability time-scales in the optical band 
and the near UV.

\item Once the variations are clearly detected, finding the structure function
eventually in different colours will be of extreme importance for a successful 
modelling of the variability.

\end{enumerate}

\section*{Acknowledgments}
We are grateful to the anonymous referee for his/hers constructive criticism, 
that helped to improve much this paper. CCD ST-8 at the Observatory 
of Belogradchik is provided by Alexander von Humboldt foundation, Germany. 
We thank Prof.~I.~Papamastorakis, Director of the Skinakas Observatory, 
and Dr.~I.~Papadakis for the allocated telescope time. 
The Skinakas Observatory is a collaborative project of the University of Crete, 
the Foundation for Research and Technology -- Hellas, and the Max-Planck-Institut 
f\"ur Extraterrestrische Physik.

\bsp

\label{lastpage}

\end{document}